\documentclass[aps,jcp,preprint,superscriptaddress,letterpaper]{revtex4}

\usepackage[utf8]{inputenc}

\usepackage{amsmath, amssymb}
\usepackage{tabularx,booktabs,array,dcolumn}
\usepackage{setspace}
\usepackage{graphicx,psfrag,subfigure}
\usepackage{xcolor}

\usepackage{marvosym}

\usepackage[all]{xy}

\usepackage{tikz}
\usetikzlibrary{calc}
\usetikzlibrary{patterns}

\usepackage{threeparttable}
\usepackage{rotating}
\usepackage{multirow}

\newcommand{\mx}[1]{\boldsymbol{#1}}
\newcommand{\bos}[1]{\boldsymbol{#1}}
\newcommand{\mr}[1]{\mathrm{#1}}

\newcommand{\pd}[2]{\frac{\partial #1}{\partial #2}}
\newcommand{\cm}{cm$^{-1}$}

\def\mel{m_\mr{e}}
\def\mpr{m_\mr{p}}
\def\iim{\text{i}}

\def\ra0{\mx{a}}

\def\som{Supplementary Material}

\def\hel{\hat{H}_\text{el}}
\def\nel{N_\text{el}}
\def\nnuc{N}

\def\Sgp{$\Sigma_\text{g}^+$}
\def\Sup{$\Sigma_\text{u}^+$}

\def\epsi{\varepsilon}
\def\hel{\hat{H}_{\text{el}}}

\def\X1Sgp{X\ ^1\Sigma_\mr{g}^+}
\def\Sgp1{^1\Sigma_\mr{g}^+}

\def\Sup1{^1\Sigma_\mr{u}^+}
\def\b3Sup{b\ ^3\Sigma_\mr{u}^+}
\def\a3Sgp{a\ ^3\Sigma_\mr{g}^+}

\def\B1Sup{B\ ^1\Sigma_\mr{u}^+}
\def\Bp1Sup{B'\ ^1\Sigma_\mr{u}^+}
\def\Pu1{^1\Pi_\mr{u}}
\def\C1Pu{C\ ^1\Pi_\mr{u}}

\def\HHbar{H\bar{H}}

\def\e3Sup{e\ ^3\Sigma_\mr{u}^+}
\def\c3Pup{c\ ^3\Pi_\mr{u}^+}

\def\Or{\mathcal{O}}

\def\EF{$EF$}
\def\GK{$GK$}
\def\HH{$H\bar{H}$}
\def\masscorr{\mathcal{A}}

\def\som{Supplementary Material}

\begin{document}

\title{%
Non-adiabatic mass correction for
excited states of molecular hydrogen: 
improvement for the outer-well $H\bar{H}\ ^1\Sigma_\mathrm{g}^+$ term values
}

\author{Dávid Ferenc and Edit Mátyus}
\email{matyus@chem.elte.hu}

\affiliation{Institute of Chemistry, ELTE, Eötvös Loránd University, Pázmány Péter sétány 1/A, Budapest, H-1117, Hungary}

\date{\today}

\begin{abstract}
\noindent %
The mass-correction function is evaluated for selected excited states
of the hydrogen molecule within a single-state non-adiabatic treatment. 
Its qualitative features are studied under the avoided
crossing of the \EF\ with the \GK\ state and also for the outer well of the \HH\ state.
For the \HH\ state, 
a negative mass correction is obtained for 
the vibrational motion near the outer minimum, 
which accounts for most of the deviation
between experiment and earlier theoretical work. \\


\end{abstract}

\maketitle

\section{Introduction}
\begin{figure}
  \includegraphics[scale=1.27]{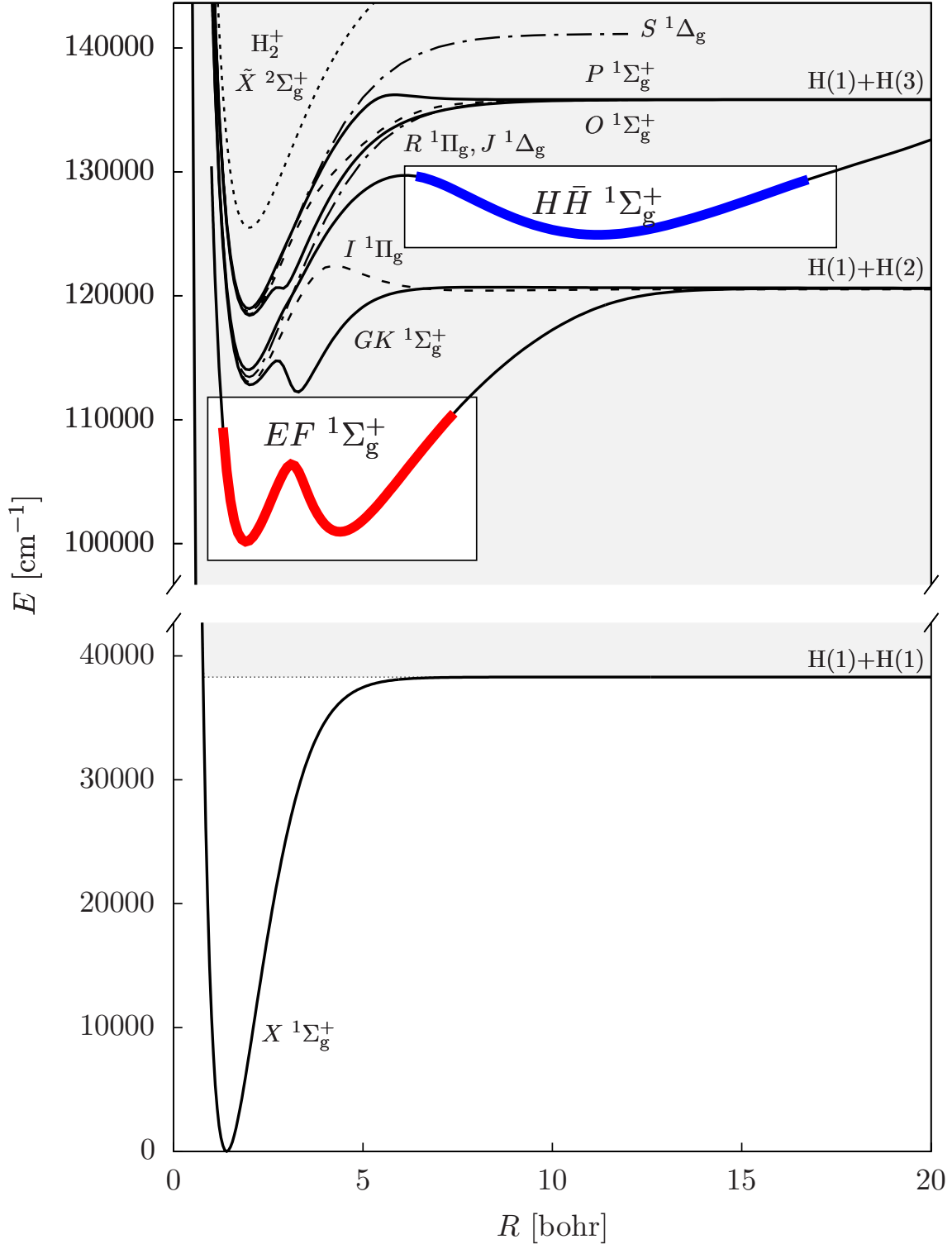}
  \caption{%
    Single-state non-adiabatic treatment for the 
    lower-energy part of the $EF\ ^1\Sigma_\text{g}^+$ electronic state, highlighted in red.     
    Results are obtained within the single-state
    non-adiabatic framework also for 
    the outer well of the $H\bar{H}\ ^1\Sigma_\text{g}^+$ electronic state, which is highlighted in blue
    (see also Table~\ref{tab:HH}).
    The Born--Oppenheimer potential energy curves shown in the figure were compiled from 
    Refs.~\cite{Wo95pi,Wo95delta,WoDr94,WoSiDa98,BeMe16}. 
    \label{fig:bopes}
  }
\end{figure}

%
\noindent This work represents the first steps towards a fully coupled non-adiabatic calculation of 
the \EF--\GK--\HH-etc. singlet-gerade manifold of H$_2$ including 
the formerly neglected mass-correction terms 
which appear in the multi-state, effective non-adiabatic Hamiltonian recently formulated 
\cite{MaTe19}.
Relying on the condition of adiabatic perturbation theory \cite{TeBook03}
that the electronic band must be separated from the rest of the electronic spectrum 
by a finite gap over the relevant dynamical range, 
already a single-state treatment delivers insight 
into the extremely rich non-adiabatic dynamics of electronically excited hydrogen.
Motivated by these ideas
and 
after careful inspection of the singlet-gerade manifold,
we have selected the lower-energy region of the $EF\ ^1\Sigma_\text{g}^+$ and 
the outer well of the $H\bar{H}\ ^1\Sigma_\text{g}^+$ state (Figure~\ref{fig:bopes}), 
often labelled with $\bar{H}$, for a single-state non-adiabatic study. 
After a short summary of the theoretical and computational details, 
we present the non-adiabatic mass curves 
and discuss them in relation with earlier theoretical and experimental work.

\section{Theoretical  and computational details}
\subsection{Summary of the theoretical background}
\noindent %
Let us start with the electronic Schr\"odinger equation, 
\begin{align}
  \hat{H}_\text{el}\psi_a = E_a \psi_a
  \label{eq:elsch}
\end{align}
including the electronic Hamiltonian (in Hartree atomic units), 
\begin{align}
  \hel
  &=
  -\frac{1}{2}\sum_{i=1}^{\nel} \Delta_{\mx{r}_i}
  + 
  \sum_{i=1}^{\nel} \sum_{j>i}^{\nel}
    \frac{1}{|\bos{r}_i-\bos{r}_j|}
  -
  \sum_{i=1}^{\nel} \sum_{k=1}^{\nnuc}
    \frac{Z_k}{|\bos{r}_i-\bos{R}_k|}
  +
  \sum_{k=1}^{\nnuc}\sum_{l>k}^{\nnuc}
    \frac{Z_k Z_l}{|\bos{R}_k-\bos{R}_l|} \, 
    \label{eq:elham}
\end{align}
with the $\mx{r}_i$ electronic and the $\mx{R}_k$ nuclear coordinates
and $Z_k$ nuclear charges.
In order to approximate the rovibronic energies of the full, electron-nucleus Schrödinger equation
accurately, it is necessary to go beyond the Born--Oppenheimer (BO) approximation.
In the present work, we explore the selected states within a single-state non-adiabatic treatment, 
using the second-order, effective Hamiltonian which had been formulated and reformulated 
in different contexts \cite{HeAs66,HeOg98,Sch01H2p,PaKo09,prx17} and 
most recently reproduced as a special case of the multi-state effective Hamiltonian \cite{MaTe19}.  
The single-state Hamiltonian has already been used for the ground electronic state of 
several diatomic molecules \cite{BuMcMo77,PaKo08,Ma18nonad,Ma18he2p,PrCeJeSz17},
in an approximate treatment of the water molecule \cite{Sch01H2O}, 
and in example single-point computations
for polyatomics \cite{prx17} (for a detailed reference list see Ref.~\cite{Ma18nonad}).

The second-order or $\epsi^2$-, effective Hamiltonian, 
for the quantum nuclear motion over a selected `$a$' electronic state is 
\begin{align}
  \hat{H}^{(2)}_{aa}
  =
  \frac{1}{2}
  \sum_{i=1}^{3\nnuc}
  \sum_{j=1}^{3\nnuc} 
    (-\iim \epsi \partial_{i})
    (\delta_{ij} - \epsi^2 M_{aa,ij})
    (-\iim \epsi \partial_{j})
  +
  \epsi^2 U_a  
  +
  E_a \; ,
  \label{eq:secHam}
\end{align}
where $\epsi^2$ is the electron-to-nucleus mass ratio,
and in particular, for the H$_2$ molecule in atomic units, it is $\epsi^2=1/\mpr$.
$\partial_i=\partial / \partial R_{k\alpha}$ with
$i=3(k-1)+\alpha$ ($k=1,\ldots,N$, $\alpha=1(x),2(y),3(z)$) 
is the partial derivative with respect to the nuclear coordinates.

Besides, the nuclear kinetic energy (with constant mass) 
and the $E_a$ electronic energy, 
$\hat{H}^{(2)}_{aa}$ contains the diagonal Born--Oppenheimer correction (DBOC),
\begin{align}
 \epsi^2 U_a 
 =
 \epsi^2
 \frac{1}{2}
 \sum_{i=1}^{3\nnuc}
 \langle
   \partial_{i} \psi_a |
   \partial_{i} \psi_a
 \rangle
 \label{eq:dboc} 
\end{align}
and the mass correction tensor, 
\begin{align}
  \epsi^2
  M_{aa,ij}
  &= \epsi^2 2 \langle \partial_j \psi_a | \mathcal{R}_a | \partial_i\psi_a \rangle \nonumber \\
  &= 
  \epsi^2 2 \langle \partial_j \psi_a | %
  (\hat{H}_\text{el}-E_\text{a})^{-1}(1-P_\text{a}) | \partial_i\psi_a \rangle\ \, ,
  \quad 
  i,j=1,\ldots 3 {\nnuc} \; ,
  \label{eq:nadcorr}
\end{align}
where the electronic energy, $E_a$, and the electronic wave function, $\psi_a$,
is obtained from solving the electronic Schrödinger equation, Eq.~(\ref{eq:elsch}).
For a better understanding of the numerical results, 
it will be important to remember the appearance of the reduced resolvent, 
$\mathcal{R}_a = (H_\text{el}-E_a)^{-1}(1-P_\text{a})$
with $P_{a}=|\psi_a\rangle \langle \psi_a|$ in the expression of the mass correction
tensor. The `effect' of the outlying electronic states
on the quantum nuclear dynamics is accounted for through this reduced resolvent.
Also note that the term containing $M_{aa,ij}$ in Eq.~(\ref{eq:secHam})
is indeed $\Or(\epsi^2)$, since 
we do not assume small nuclear momenta, and hence
the action of $\hat{p}_i = -\iim\epsi\partial_{i}$  
on the nuclear wave function creates an $\Or(1)$ contribution instead of $\Or(\epsi)$ 
(for more details, see for example Ref.~\cite{MaTe19}).

The general transformation of the second-order non-adiabatic kinetic energy operator
for an $\nnuc$-atomic molecule---first term in Eq.~(\ref{eq:secHam})---to curvilinear coordinates 
was worked out in Ref.~\cite{Ma18nonad}. The special transformation for a diatomic molecule 
described with spherical polar coordinates, $(\rho,\Omega)$,
which we are using in the present work, had been formulated earlier 
\cite{HeAs66,BuMo77,BuMcMo77,PaKo09}.
Hence, the effective, non-adiabatic, single-state Schrödinger equation
for the hydrogen molecule with $J$ rotational angular momentum quantum number is
\begin{align}
&\left[%
  -\pd{}{\rho} \frac{1}{\mpr} \left(1 - \frac{(\masscorr_{a})^{\rho}\ _\rho}{\mpr}\right) \pd{}{\rho}
  +\frac{J(J+1)}{\mpr\ \rho^2} \left(1 - \frac{(\masscorr_{a})^{\Omega}\ _\Omega}{\mpr}\right)
  +\frac{1}{m_\text{p}} U_a(\rho)+V_a(\rho)
\right]
\phi_{J}(\rho)
\nonumber \\
&
\quad\quad\quad=
E_J\phi_J(\rho) 
\label{eq:nucHamdiat}
\end{align}
with the volume element  $\text{d}\rho$.
The $(\masscorr_{a})^{\rho}\ _\rho$ and $(\masscorr_{a})^{\Omega}\ _\Omega$
(coordinate-dependent) coefficients originate from the mass-correction tensor, Eq.~(\ref{eq:nadcorr}), 
and the coordinate transformation rule from Cartesian coordinates to curvilinear coordinates 
(see Ref.~\cite{Ma18nonad,Ma18he2p}).
Since, $1/(1+y)\approx 1-y$ for small $y$, we may write 
$1/\mpr (1-a/\mpr) \approx 1/\mpr \cdot 1/(1+a/\mpr) = 1/(\mpr + a)$,
and in this sense, $(\masscorr_{a})^{\rho}\ _\rho$ and $(\masscorr_{a})^{\Omega}\ _\Omega$
can be interpreted as correction to the proton mass for
the vibrational and rotational motion, or in short, vibrational and rotational mass correction, 
respectively.


\subsection{Computational details}
We used the QUANTEN computer program \cite{MaRe12,Ma19review,Ma18nonad,Ma18he2p} 
to accurately solve the electronic Schrödinger equation, Eq.~(\ref{eq:elsch}),
for the second, third, and fourth 
singlet gerade electronic states---$EF$, $GK$, and $H\bar{H}$ $^1\Sigma_\text{g}^+$, 
respectively---of the hydrogen molecule using floating, explicitly correlated Gaussian functions
 as spatial functions
\begin{align}
  f(\mx{r};\mx{A},\mx{s})
  =
  \exp\left[%
    -\frac{1}{2} 
    (\mx{r}-\mx{s})^\text{T} 
    (\mx{A}\otimes \mx{I}_3)
    (\mx{r}-\mx{s})
  \right] \; .
\end{align}
The $\mx{A}\in\mathbb{R}^{2\times 2}$ and $\mx{s}\in\mathbb{R}^{2\cdot 3}$
parameters were
optimized variationally for each basis function at every nuclear configuration, $\mx{R}_{k}\ (k=1,2)$,
over a fine grid of the $\rho=|\mx{R}_1-\mx{R}_2|$ nuclear separation.
For the computational details of the wave function derivatives and the mass correction functions
in curvilinear coordinates see Refs.~\cite{Ma18he2p,Ma18nonad}.

The electronic states we study in this work
and the corresponding DBOCs and relativistic corrections 
have already been computed accurately by Wolniewicz \cite{Wo98rel}
for several proton-proton distances. We have repeated these computations
to check the accuracy of the data, and to improve
upon it where it was necessary \emph{(vide infra).}
In addition, we have computed the non-adiabatic mass-correction functions in the present work.

The effective nuclear Schrödinger equation for the diatom, Eq.~(\ref{eq:nucHamdiat}), was solved
using the discrete variable representation (DVR) \cite{LiCa00} 
and associated Laguerre polynomials, $L_n^{(\alpha)}$
with $\alpha=2$ for the vibrational ($\rho$) degree of freedom.
The outer-well $\bar{H}$ states were computed by scaling the DVR points 
to the $[R_\text{min},R_\text{max}]=[6,20]$~bohr interval.

\section{Results and discussion}
\begin{figure}
  \includegraphics[scale=0.85]{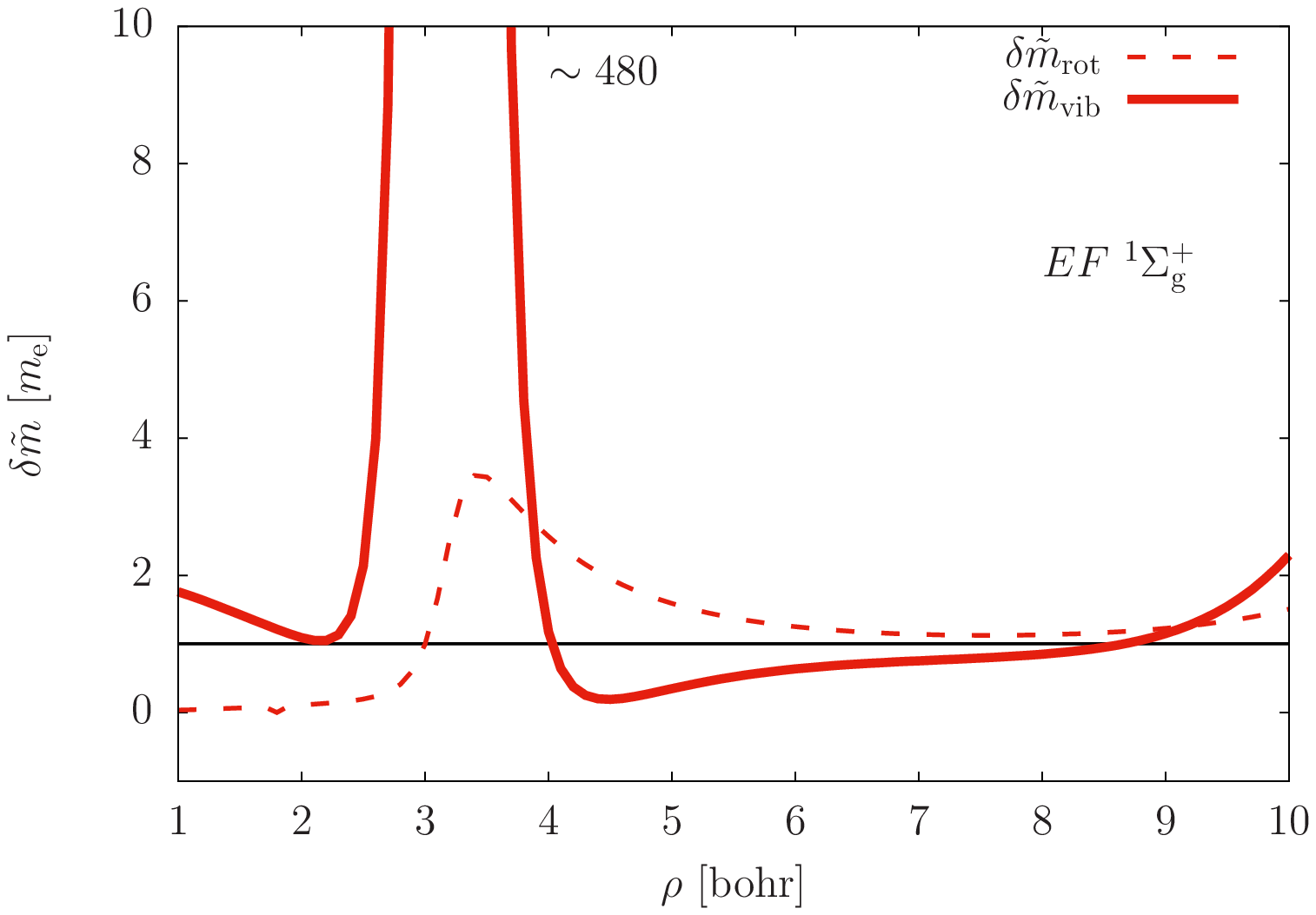}
  \caption{%
  Non-adiabatic mass correction functions to the rotational and the vibrational
  degrees of freedom, $\delta \tilde{m}_\text{rot}$ and $\delta \tilde{m}_\text{vib}$,
  computed for 
  the $EF\ ^1\Sigma_\text{g}^+$ electronic state of the hydrogen molecule \cite{Ma18nonad,Ma18he2p}.
  (The thin, solid black line indicates the mass of the electron, which together with the proton
  mass gives the atomic mass.)
  \label{fig:masscorrEF}
  }
\end{figure}

\noindent The non-adiabatic correction to the vibrational and rotational
mass of the proton in the $EF$ and $\HHbar$ $^1\Sigma_\text{g}^+$ electronic states
of H$_2$ are shown in Figures~\ref{fig:masscorrEF} and \ref{fig:masscorrHH}. 
These numerical examples highlight qualitative features of the mass-correction
functions. The sign and the amplitude of the correction can be understood by 
remembering that the mass-correction tensor 
contains the $\mathcal{R}_\text{a}$ reduced resolvent, Eq.~(\ref{eq:nadcorr}).

\vspace{0.5cm}
\paragraph{The $EF$ state}
Concerning the $EF\ ^1\Sigma_\text{g}^+$ state of H$_2$ 
below ca.~110\,000~\cm\  (below the $GK$ minima, Figure~\ref{fig:bopes}),
the effective vibrational mass of the proton
becomes gigantic under the avoided crossing with the $GK\ ^1\Sigma_\text{g}^+$ curve.
The large correction value, $\delta m_\text{vib}^{(EF)}=480\ \mel$ near $R=3$~bohr, 
which should be compared with the ca. $1836\ \mel$ mass of the proton
\footnote{In the computations we used the precise values of the CODATA14 constants
and conversion factors, {h}ttp://physics.nist.gov/cuu/Constants (last accessed on 3 May 2019).} , 
indicates that it will be necessary to go beyond the second-order, single-state non-adiabatic treatment
to achieve spectroscopic accuracy.
For this purpose, one can either consider using higher-order corrections---the third-order correction formulae can be 
found in Ref.~\cite{MaTe19}---, or including explicit non-adiabatic coupling 
with the near-lying perturber state(s), in this case $GK$ (and probably other states),
and to use the effective non-adiabatic Hamiltonian of Ref.~\cite{MaTe19} for a multi-dimensional 
electronic subspace.
Note that in the single-state treatment only the $EF$ state is projected out
from the resolvent, Eq.~(\ref{eq:nadcorr}),
whereas in a multi-state treatment the full explicitly coupled subspace will be projected out, 
which will result in smaller corrections from
electronic states better separated in energy.

With these observations in mind, we have nevertheless checked the rotation-vibration term values obtained within 
the second-order, single-state non-adiabatic model. We have found that the $EF$ vibrational
term values obtained with the effective masses (Figure~\ref{fig:masscorrEF}) were closer to the experimental values
than the constant-mass adiabatic description (using either the nuclear mass of 
the proton or the atomic mass of hydrogen,
which is commonly used as an `empirical' means of modeling non-adiabatic effects).
The ca.~30--35~\cm\ root-mean-square deviation 
of the adiabatic energies from experiment 
was reduced to 10~\cm\ 
when the rigorous non-adiabatic vibrational functions were used
instead of the constant (nuclear or atomic) mass.
More detailed numerical results will be obtained within a coupled-state non-adiabatic treatment
in future work.

\vspace{0.5cm}
\paragraph{The $H\bar{H}$ state}
Next, we have studied the $H\bar{H}$ state, 
for which already a single-state model turns out to be useful for spectroscopic purposes, 
at least for the outer-well states.
The inner well of the $H\bar{H}$ potential energy curve (PEC) gets close to several other PECs, 
and for this reason a single-state treatment is not appropriate there.
At the same time, most of the outer-well state energies (below the barrier) can be accurately
computed without considering delocalization to the inner well. 
This behaviour was pointed out already several times in the literature
\cite{Wo98rel,Wo98HH,ReHoUbWo99,AnEl04,RoYoOgTs06}, 
and we have also checked it for every rovibrational state
by solving the rovibrational Schr\"odinger equation with different 
$[R_\text{min},R_\text{max}]$ intervals. 
In particular, we obtained the (adiabatic) inner-well state energies 
(below the barrier) with an accuracy better than $0.01$~\cm\ 
even if we used the restricted, $[R_\text{min},R_\text{max}]=[6,20]$~bohr, interval.
This behavior was observed either with using constant (\emph{e.g.,} nuclear or atomic) or 
coordinate-dependent, non-adiabatic masses. 
The few exceptions (states nearer the top of the barrier 
which separates the inner and the outer wells)
are shown in gray in Table~\ref{tab:HH}.

The experimental term values for the outer-well rotation-vibration states of 
the $H\bar{H}$ electronic state
were first reported in 1997 \cite{ReHoUb97} and also later in 1999 together with an improved
theoretical treatment \cite{ReHoUbWo99,Wo98HH}. 
The computations were carried out on an accurate, adiabatic PEC
including relativistic corrections and were appended also with an estimate
for the quantum electrodynamics (QED) effects \cite{Wo98rel,ReHoUbWo99}.
The resulting term values were in a ca.~1~\cm\ (dis)agreement with experiment 
(of ca. 0.04~\cm\ uncertainty), which was attributed to the neglect of
non-adiabatic effects.

We have repeated the rovibrational computations using the potential energy, 
diagonal Born--Oppenheimer correction, and relativistic correction curves computed 
and the QED correction estimated by Wolniewicz \cite{Wo98rel}, 
but we used the non-adiabatic mass correction functions for the rotational and the vibrational
degrees of freedom computed in the present work (Figure~\ref{fig:masscorrHH}). 
We obtained a somewhat better agreement 
with the experimental results, the 1--1.2~\cm\ deviation of theory and experiment 
of Ref.~\cite{ReHoUbWo99} was reduced to 0.3--0.4~\cm\ 
(the computed values are larger than the experimental ones).

In order to identify the origin of the remaining discrepancy, we refined
the potential energy curve using the QUANTEN program, 
which resulted in a few tenths of \cm\ reduction for $R>10$~bohr 
(the improved electronic energies are deposited in the \som~\cite{som}).
Next, we have checked the accuracy of the earlier relativistic corrections and found them to be sufficient
for the present purposes.
We have also explicitly evaluated 
the leading-order QED correction (see for example, Eq.~(3) of Ref.~\cite{FeMa19a}), 
instead of approximating it with the QED
correction value of H$^-$ proposed by Wolniewicz \cite{Wo98rel}.
For this purpose, we 
used the one- and two-electron Darwin integrals already available from 
the relativistic computations \cite{Wo98rel}
and approximated the non-relativistic Bethe-logarithm with $\ln k_0\approx 3$ 
based on its value for the ground state of the hydrogen atom 
(remember the strong H$^-$+H$^+$ ion-pair character 
of the outer well and the observation that $\ln k_0$ 
is not very sensitive to the number of electrons \cite{Ko18H2p}). 
We also computed the Araki--Sucher term for the $H\bar{H}$ state in the present work,
although it gives an almost negligible contribution at the current level of precision. 
Based on these computations, 
the radiative correction curve takes values between
0.27 and 0.29~\cm\ over the outer well of $H\bar{H}$,
and thus we confirm the earlier estimate using the H$^-$ value \cite{Wo98rel}.

As a summary, 
we collect in Table~\ref{tab:HH}
the best rovibrational term values (`$T_\text{nad}$' column)
and their deviation from experiment (`nad' column) resulting 
from the computations carried out within the present work. 
Inclusion of the non-adiabatic
masses in the rovibrational treatment and further refinement of the potential energy curve
reduces the earlier ca.~1--1.5~\cm\ deviation to ca.~0.1--0.2~\cm. 
The $v\geq 14$ states are shown in grey in the table, 
because for these states tunneling to the inner well 
has an important effect on the energy, 
and an `isolated' outer-well treatment is not sufficient for these states.
We also note that 
the experimental term values 
for the $v=2$ states with $N=0,1,\ldots,4$ and the $v=2$ and 3 states for $N=5$
are an order-of-magnitude less accurate than for
the other states \cite{ReHoUbWo99,ReHoUb97}. 

In the table, we also compare with experiment 
the adiabatic energies 
(a) computed rigorously with the nuclear masses 
(`ad$_\text{p}$' column)---these values are almost identical with the values in Table~VI of Ref.~\cite{ReHoUbWo99}---, 
and 
(b) with the hydrogenic atomic mass (`ad$_\text{H}$' column), 
which is often used to capture some non-adiabatic effects
in the spectrum. 
In the present case, this atomic-mass model does not perform 
well, which can be understood by noticing that the rigorous non-adiabatic (vibrational) correction 
to the nuclear mass is negative over most of the outer well.

Finally, we mention that Andersson and Elander \cite{AnEl04},
by extending earlier work of Yu and Dressler \cite{YuDr94},
solved the coupled-state equations, including the coupling of the six lowest-energy $^1\Sigma_\text{g}^+$ 
states, and studied also the outer-well region of $H\bar{H}$. They found
that it was necessary to include all six $^1\Sigma_\text{g}^+$ states to converge 
the $\bar{H}$ vibrational energies better than 0.1~\cm\, whereas the 15th and 16th vibrational states
($v=14$ and 15 in Table~\ref{tab:HH}) changed by 0.12 and 24.11~\cm\ between the five- 
and six-state treatment. Although their computed values are off by 10--35~\cm\ from experiment, 
probably due to the fact that they used less accurate potential energy curves, 
their results seem to underline the general observation that
the many-state Born--Oppenheimer (BO) expansion converges 
relatively slowly with respect to the number of electronic states.

\section{Summary and conclusions}
\noindent 
Due to the slow convergence of the Born--Oppenheimer (BO) expansion with respect to the number of electronic states,
it is important to think about the truncation error 
when one is aiming to compute highly accurate molecular rovibrational (rovibronic) energies.
Direct truncation introduces an error of $\Or(\epsi)$ in the rovibronic 
energies, where $\epsi=(\mel/m_\text{nuc})^{1/2}$ is the 
square root of the electron-to-nucleus mass ratio \cite{MaTe19,TeBook03,PaSpTe07}.
This truncation error can be made lower order in $\epsi$ by 
using adiabatic perturbation theory \cite{TeBook03,PaSpTe07}.
For an isolated electronic state, the first-order corrections can be made to vanish.
The second-order non-adiabatic effective Hamiltonian, used in the present work,
reproduces eigenvalues of the full electron-nucleus Hamiltonian with an error of $\Or(\epsi^3)$,
but it contains corrections both to the potential energy 
as well as to the kinetic energy of the atomic nuclei \cite{MaTe19},
which gives rise to effective coordinate-dependent masses to the different types of motions.

\begin{figure}
  \includegraphics[scale=0.85]{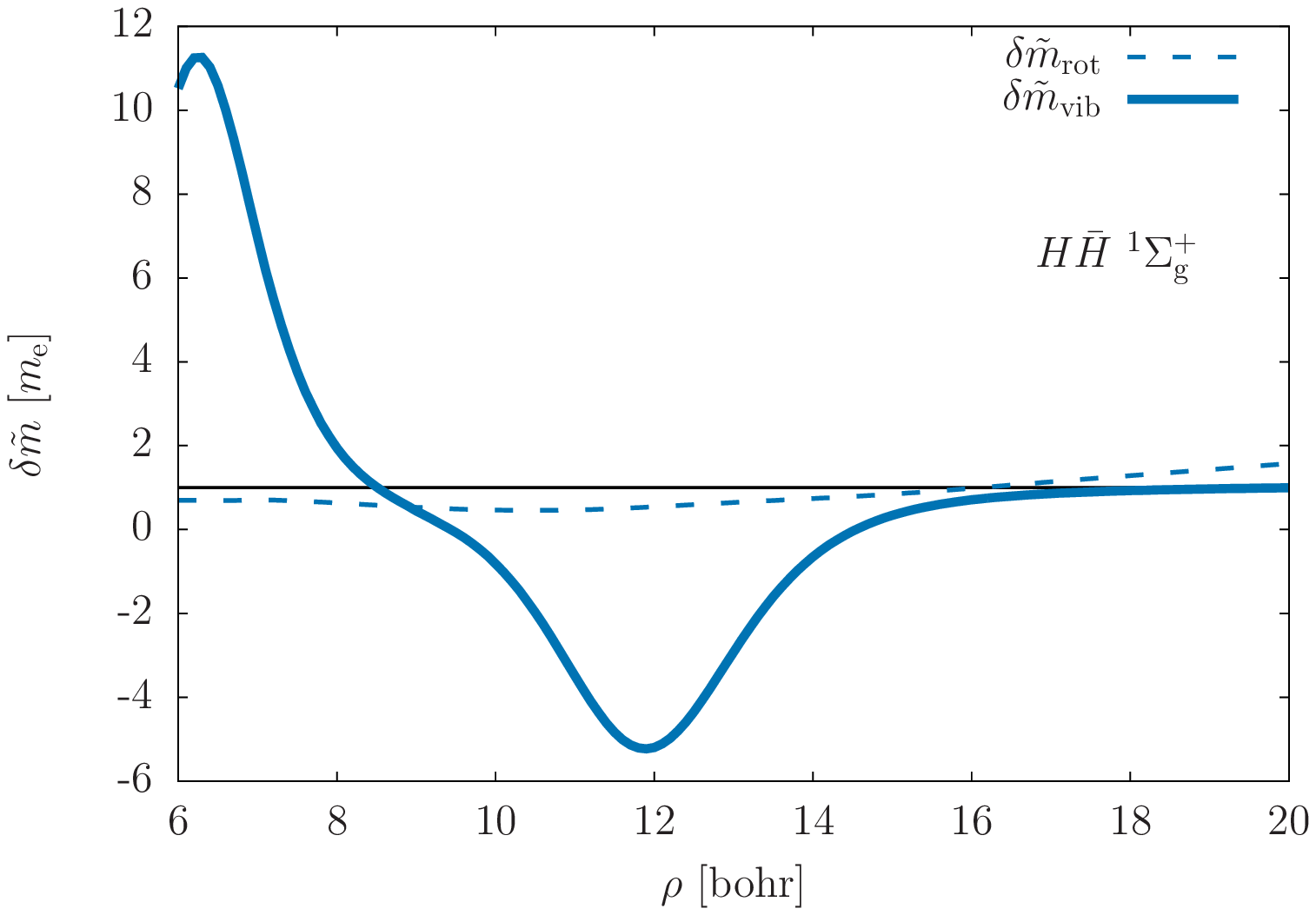} 
  \caption{%
  Non-adiabatic mass correction functions to the rotational and the vibrational
  degrees of freedom, $\delta \tilde{m}_\text{rot}$ and $\delta \tilde{m}_\text{vib}$, 
  computed for 
  the $H\bar{H}\ ^1\Sigma_\text{g}^+$ electronic state of the hydrogen molecule \cite{Ma18nonad,Ma18he2p}.
  (The thin, solid black line indicates the mass of the electron, which together with the proton
  mass gives the atomic mass.)
  \label{fig:masscorrHH}
  }
\end{figure}

In particular, we have found a non-trivial, negative mass-correction to the vibrational mass of the proton
in the outer well of the $H\bar{H}\ ^1\Sigma_\text{g}^+$ electronic
state. This negative value, \emph{i.e.,} an effective vibrational mass smaller than the nuclear mass,
is dominated by the interaction with the H(1)+H(2) dissociation channel 
to which $H\bar{H}$ gets close near its outer
minimum. Of course, the precise value of the mass correction is the result of an interplay of
the interaction of the nuclear dynamics on $H\bar{H}$ with all the other (discrete and continuous) electronic states.
It is interesting to note that, whereas the vibrational mass shows this special behaviour
for $H\bar{H}$, the non-adiabatic value of the rotational mass remains 
close to the \emph{atomic mass} of the hydrogen (proton plus electron, see Figure~\ref{fig:masscorrHH}).
Due to these properties, $H\bar{H}$ makes a counter-example to the simple, empirical recipe 
according to which small non-adiabatic effects can be `approximately modeled' by using 
(near) the atomic mass value for vibrations and the nuclear mass for rotations \cite{PoTe99,DiMoAl13,MaSzCs14}.
In the case of the outer well of $H\bar{H}$, the vibrational mass is 
better approximated by the nuclear mass,
while the rotational mass equals the atomic mass to a good approximation.
Using the rigorous non-adiabatic, mass-correction functions computed in the present work, 
the non-adiabatic rovibrational energies 
are ca.~1~\cm\ (2~\cm) larger than the energies obtained using the nuclear (atomic) mass. 
This, together with the relativistic and radiative corrections as well as with a minor, 0.1--0.2~\cm\ 
improvement for the outer-well electronic energies, allows us to achieve a 0.1--0.2~\cm\ agreement,
an order of magnitude better than earlier theory,
with experiment \cite{ReHoUb97,ReHoUbWo99}.

All in all, we have demonstrated that small, non-adiabatic corrections in 
the (high-resolution) spectrum can be
efficiently described using the effective non-adiabatic Hamiltonian which accounts
for the truncation error in the electronic space perturbatively.
For the particular case of the outer well of the $H\bar{H}\ ^1\Sigma_\text{g}^+$ electronic state,
the discrepancy of earlier theoretical work with experiment can be accounted for
by a non-trivial decrease in the effective, non-adiabatic vibrational mass of the protons 
as they pass along near-lying electronic states.

\vspace{0.5cm}
\paragraph*{Acknowledgement}
\noindent %
We acknowledge financial support from a PROMYS Grant (no. IZ11Z0\_166525)  
of the Swiss National Science Foundation.
DF thanks a doctoral scholarship from
the New National Excellence Program of 
the Ministry of Human Capacities of Hungary
(ÚNKP-18-3-II-ELTE-133). 
EM thanks ETH~Z\"urich for a visiting professorship during 2019 and 
the Laboratory of Physical Chemistry for their hospitality, where
part of this work has been completed.


\clearpage
\begin{table}
\caption{%
  Term values and deviation from experiment, in \cm,
  for the outer-well rovibrational states of 
  the $H\bar{H}\ ^1\Sigma_\text{g}^+$ electronic state
  of the hydrogen molecule.
\label{tab:HH}}
\scalebox{1.}{%
\begin{tabular}{@{}cc c@{\ \ }r@{\ }r@{\ }r@{}r@{\ \ } c@{\ \ }r@{\ }r@{\ }r@{}r@{\ \ } c@{\ \ }r@{\ }r@{\ }r @{}}
  \cline{1-16} \\[-0.38cm]
  \cline{1-16} \\[-0.25cm]  
  && 
  \multicolumn{4}{c}{$J=0$} &&
  \multicolumn{4}{c}{$J=1$} && 
  \multicolumn{4}{c}{$J=2$} \\
  \cline{3-6} 
  \cline{8-11}
  \cline{13-16} \\[-0.25cm]
  &&
  & \multicolumn{3}{c}{$T_\text{obs}-T_\text{calc}$$^\text{a}$} && 
  & \multicolumn{3}{c}{$T_\text{obs}-T_\text{calc}$$^\text{a}$} && 
  & \multicolumn{3}{c}{$T_\text{obs}-T_\text{calc}$$^\text{a}$}  \\ 
  \cline{4-6} 
  \cline{9-11}
  \cline{14-16} \\[-0.25cm]
  $v$ &&
  $T_\text{nad}$$^\text{b}$ & nad$^\text{c}$ & ad$_\text{H}$$^\text{d}$ & ad$_\text{p}$$^\text{e}$ && 
  $T_\text{nad}$$^\text{b}$ & nad$^\text{c}$ & ad$_\text{H}$$^\text{d}$ & ad$_\text{p}$$^\text{e}$ && 
  $T_\text{nad}$$^\text{b}$ & nad$^\text{c}$ & ad$_\text{H}$$^\text{d}$ & ad$_\text{p}$$^\text{e}$   \\ 
  \cline{1-16} \\[-0.25cm]
  %
0 && 
122883.4 &  & n.a.$^\mathrm{f}$ & &&  
122885.3 &  & n.a.$^\mathrm{f}$ & &&  
122889.1 &  & n.a.$^\mathrm{f}$ & \\
1 && 
123234.5 & & n.a.$^\mathrm{f}$ & &&  
123236.4 & & n.a.$^\mathrm{f}$ & &&  
123240.2 & & n.a.$^\mathrm{f}$ & \\
2 && 
123575.8 & $ 0.1 $$^\mathrm{g}$ & $ 1.1 $$^\mathrm{g}$ & $ 0.9 $$^\mathrm{g}$ &&  
123577.7 & $ 0.0 $$^\mathrm{g}$ & $ 1.0 $$^\mathrm{g}$ & $ 0.8 $$^\mathrm{g}$ &&  
123581.5 & $ 0.2 $$^\mathrm{g}$ & $ 1.1 $$^\mathrm{g}$ & $ 0.9 $$^\mathrm{g}$ \\
3 && 123907.5 & $ -0.2 $ & $ 1.1 $ & $ 0.8 $ &&  123909.4 & $ -0.2 $ & $ 1.1 $ & $ 0.8 $ &&  123913.3 & $ -0.2 $ & $ 1.1 $ & $ 0.8$ \\
4 && 124229.9 & $ -0.2 $ & $ 1.3 $ & $ 0.9 $ &&  124231.8 & $ -0.2 $ & $ 1.3 $ & $ 0.9 $ &&  124235.7 & $ -0.3 $ & $ 1.2 $ & $ 0.8$ \\
5 && 124543.0 & $ -0.3 $ & $ 1.4 $ & $ 0.9 $ &&  124544.9 & $ -0.3 $ & $ 1.4 $ & $ 0.9 $ &&  124548.8 & $ -0.3 $ & $ 1.4 $ & $ 0.9$ \\
6 && 124847.2 & $ -0.2 $ & $ 1.6 $ & $ 1.1 $ &&  124849.1 & $ -0.2 $ & $ 1.6 $ & $ 1.0 $ &&  124853.0 & $ -0.2 $ & $ 1.6 $ & $ 1.0$ \\
7 && 125142.5 & $ -0.2 $ & $ 1.7 $ & $ 1.1 $ &&  125144.5 & $ -0.2 $ & $ 1.7 $ & $ 1.1 $ &&  125148.4 & $ -0.2 $ & $ 1.7 $ & $ 1.1$ \\
8 && 125429.3 & $ -0.2 $ & $ 1.8 $ & $ 1.2 $ &&  125431.3 & $ -0.1 $ & $ 1.8 $ & $ 1.2 $ &&  125435.1 & $ -0.2 $ & $ 1.8 $ & $ 1.1$ \\
9 && 125707.6 & $ -0.1 $ & $ 1.9 $ & $ 1.2 $ &&  125709.6 & $ -0.1 $ & $ 2.0 $ & $ 1.3 $ &&  125713.5 & $ -0.1 $ & $ 1.9 $ & $ 1.2$ \\
10 && 125977.6 & $ -0.1 $ & $ 1.9 $ & $ 1.2 $ &&  125979.5 & $ -0.1 $ & $ 2.0 $ & $ 1.2 $ &&  125983.4 & $ -0.1 $ & $ 1.9 $ & $ 1.2$ \\
11 && 126239.1 & $ -0.1 $ & $ 1.9 $ & $ 1.1 $ &&  126241.1 & $ -0.1 $ & $ 2.0 $ & $ 1.2 $ &&  126245.0 & $ -0.2 $ & $ 1.9 $ & $ 1.1$ \\
12 && 126492.4 & $ -0.2 $ & $ 1.9 $ & $ 1.1 $ &&  126494.4 & $ -0.2 $ & $ 1.9 $ & $ 1.0 $ &&  126498.3 & $ -0.2 $ & $ 1.9 $ & $ 1.0$ \\
\color{gray}13$^\text{h}$ && 
\color{gray}126737.1 & \color{gray}$ -0.2 $ & \color{gray}$ 1.8 $ & \color{gray}$ 0.9 $ &&  
\color{gray}126739.1 & \color{gray}$ -0.2 $ & \color{gray}$ 1.8 $ & \color{gray}$ 0.9 $ &&  
\color{gray}126743.1 & \color{gray}$ -0.3 $ & \color{gray}$ 1.8 $ & \color{gray}$ 0.9$ \\
\color{gray}14$^\text{h}$ && 
\color{gray}126972.9 & \color{gray}$ -0.5 $ & \color{gray}$ 1.5 $ & \color{gray}$ 0.6 $ &&  
\color{gray}126975.0 & \color{gray}$ -0.5 $ & \color{gray}$ 1.5 $ & \color{gray}$ 0.6 $ &&  
\color{gray}126979.1 & \color{gray}$ -0.6 $ & \color{gray}$ 1.4 $ & \color{gray}$ 0.5$ \\
\color{gray}15$^\text{h}$ && 
\color{gray}127199.3 & \color{gray}$ -1.8 $ & \color{gray}$ 0.2 $ & \color{gray}$ -0.8 $ &&  
\color{gray}127201.4 & \color{gray}$ -1.7 $ & \color{gray}$ 0.3 $ & \color{gray}$ -0.7 $ &&  
\color{gray}127205.6 & \color{gray}$ -1.5 $ & \color{gray}$ 0.4 $ & \color{gray}$ -0.5$ \\

  \cline{1-16} \\[-0.25cm]
    && 
  \multicolumn{4}{c}{$J=3$} &&
  \multicolumn{4}{c}{$J=4$} && 
  \multicolumn{4}{c}{$J=5$} \\
  \cline{3-6} 
  \cline{8-11}
  \cline{13-16} \\[-0.25cm]
  &&
  & \multicolumn{3}{c}{$T_\text{obs}-T_\text{calc}$$^\text{a}$} && 
  & \multicolumn{3}{c}{$T_\text{obs}-T_\text{calc}$$^\text{a}$} && 
  & \multicolumn{3}{c}{$T_\text{obs}-T_\text{calc}$$^\text{a}$} \\ 
  \cline{4-6} 
  \cline{9-11}
  \cline{14-16} \\[-0.25cm]
  $v$ &&
  $T_\text{nad}$$^\text{b}$ & nad$^\text{c}$ & ad$_\text{H}$$^\text{d}$ & ad$_\text{p}$$^\text{e}$ && 
  $T_\text{nad}$$^\text{b}$ & nad$^\text{c}$ & ad$_\text{H}$$^\text{d}$ & ad$_\text{p}$$^\text{e}$ && 
  $T_\text{nad}$$^\text{b}$ & nad$^\text{c}$ & ad$_\text{H}$$^\text{d}$ & ad$_\text{p}$$^\text{e}$ \\ 
  \cline{1-16} \\[-0.25cm]  
  %
0 && 
122894.9 &  & n.a.$^\mathrm{f}$ &   &&  
122902.5 &  & n.a.$^\mathrm{f}$ &   && 
122912.0 &  & n.a.$^\mathrm{f}$ &    \\
1 && 
123246.0 &  & n.a.$^\mathrm{f}$ &   &&  
123253.6 &  & n.a.$^\mathrm{f}$ &   &&  
123263.2 &  & n.a.$^\mathrm{f}$ &   \\
2 && 
123587.3 & $ 0.0 $$^\mathrm{g}$ & $ 1.0 $$^\mathrm{g}$ & $ 0.8 $$^\mathrm{g}$ &&  
123595.0 & $ 0.4 $$^\mathrm{g}$ & $ 1.4 $$^\mathrm{g}$ & $ 1.2 $$^\mathrm{g}$ &&  
123604.6 & $ -0.3 $$^\mathrm{g}$ & $ 0.7 $$^\mathrm{g}$ & $ 0.5$$^\mathrm{g}$ \\
3 && 123919.1 & $ -0.2 $ & $ 1.0 $ & $ 0.7 $ &&  
123926.8 & $ -0.2 $ & $ 1.1 $ & $ 0.7 $ &&  
123936.4 & $ 0.0 $$^\mathrm{g}$ & $ 1.3 $$^\mathrm{g}$ & $ 0.9$$^\mathrm{g}$ \\
4 && 124241.5 & $ -0.2 $ & $ 1.2 $ & $ 0.9 $ &&  124249.2 & $ -0.3 $ & $ 1.2 $ & $ 0.8 $ &&  124258.8 & $ -0.2 $ & $ 1.3 $ & $ 0.9$ \\
5 && 124554.6 & $ -0.2 $ & $ 1.4 $ & $ 1.0 $ &&  124562.3 & $ -0.3 $ & $ 1.4 $ & $ 0.9 $ &&  124572.0 & $ -0.2 $ & $ 1.4 $ & $ 0.9$ \\
6 && 124858.8 & $ -0.2 $ & $ 1.6 $ & $ 1.1 $ &&  124866.5 & $ -0.2 $ & $ 1.6 $ & $ 1.0 $ &&  124876.2 & $ -0.2 $ & $ 1.6 $ & $ 1.0$ \\
7 && 125154.2 & $ -0.1 $ & $ 1.8 $ & $ 1.2 $ &&  125161.9 & $ -0.2 $ & $ 1.7 $ & $ 1.1 $ &&  125171.6 & $ -0.2 $ & $ 1.7 $ & $ 1.1$ \\
8 && 125441.0 & $ -0.2 $ & $ 1.8 $ & $ 1.2 $ &&  125448.7 & $ -0.2 $ & $ 1.8 $ & $ 1.1 $ &&  125458.4 & $ -0.2 $ & $ 1.8 $ & $ 1.2$ \\
9 && 125719.3 & $ -0.1 $ & $ 1.9 $ & $ 1.2 $ &&  125727.0 & $ -0.1 $ & $ 1.9 $ & $ 1.2 $ &&  125736.7 & $ -0.1 $ & $ 1.9 $ & $ 1.2$ \\
10 && 125989.3 & $ -0.1 $ & $ 2.0 $ & $ 1.2 $ &&  125997.0 & $ -0.1 $ & $ 1.9 $ & $ 1.2 $ &&  126006.8 & $ -0.1 $ & $ 1.9 $ & $ 1.2$ \\
11 && 126250.9 & $ -0.1 $ & $ 2.0 $ & $ 1.2 $ &&  126258.7 & $ -0.2 $ & $ 1.9 $ & $ 1.1 $ &&  126268.5 & $ -0.1 $ & $ 1.9 $ & $ 1.1$ \\
12 && 126504.2 & $ -0.2 $ & $ 1.9 $ & $ 1.0 $ &&  126512.1 & $ -0.2 $ & $ 1.9 $ & $ 1.0 $ &&  126522.0 & $ -0.2 $ & $ 1.9 $ & $ 1.0$ \\
\color{gray}13$^\text{h}$ && 
\color{gray}126749.1 & \color{gray}$ -0.3 $ & \color{gray}$ 1.8 $ & \color{gray}$ 0.9 $ &&  
\color{gray}126757.1 & \color{gray}$ -0.3 $ & \color{gray}$ 1.8 $ & \color{gray}$ 0.9 $ &&  
\color{gray}126767.1 & \color{gray}$ -0.3 $ & \color{gray}$ 1.8 $ & \color{gray}$ 0.9$ \\
\color{gray}14$^\text{h}$ && 
\color{gray}126985.2 & \color{gray}$ -0.4 $ & \color{gray}$ 1.6 $ & \color{gray}$ 0.7 $ &&  
\color{gray}126993.4 & \color{gray}$ -0.6 $ & \color{gray}$ 1.4 $ & \color{gray}$ 0.5 $ &&  
\color{gray}127003.6 & \color{gray}$ -0.4 $ & \color{gray}$ 1.7 $ & \color{gray}$ 0.7$ \\
\color{gray}15$^\text{h}$ && 
\color{gray}127211.9 & \color{gray}$ -1.3 $ & \color{gray}$ 0.7 $ & \color{gray}$ -0.3 $ &&  
\color{gray}127220.4 & \color{gray}$ -0.9 $ & \color{gray}$ 1.1 $ & \color{gray}$ 0.1 $ &&  
\color{gray}127230.8 & $   $ &  $  $ & $ $ \\
  \cline{1-16} \\[-0.38cm]
  \cline{1-16} \\[-0.25cm]  
\end{tabular}
}
\begin{flushleft}
 \emph{(Please find the footnotes on the next page)}
\end{flushleft}

\end{table}
\clearpage
\noindent \emph{Footnotes to Table~\ref{tab:HH}}
\begin{flushleft}
{
  $^\text{a}$ %
  Deviation of experiment and theory. 
  The $T_\text{obs}$ experimental term values were taken from Refs.~\cite{ReHoUb97,ReHoUbWo99}. \\  
  $^\text{b}$ %
  Calculated term value, $T_\text{nad}=E_\text{nad}-E_0$, referenced to 
  the ground-state energy, $E_0$ \cite{WaYa18,PuSpKoPa18}.
  $E_\text{nad}$ was obtained by solving Eq.~(\ref{eq:nucHamdiat}) 
  with the rigorous non-adiabatic masses computed 
  in the present work (Figure~\ref{fig:masscorrHH}) and using the relativistic 
  and diagonal Born--Oppenheimer corrections
  of Ref.~\cite{Wo98rel,Wo98HH}, as well as the radiative corrections and an improved PEC 
  computed in this work. \\  
  $^\text{c}$ %
  $T_\text{calc}=T_\text{nad}$. \\
  $^\text{d}$ %
  $T_\text{calc}=T_{\text{ad}_\text{H}}=E_{\text{ad}_\text{H}}-E_0$, where
  $E_{\text{ad}_\text{H}}$ was obtained as $E_\text{nad}$ but using the constant,
  atomic mass of hydrogen for $m_\text{p}$ and approximating the mass-correction functions by zero. \\
  $^\text{e}$ %
  $T_\text{calc}=T_{\text{ad}_\text{p}}=E_{\text{ad}_\text{p}}-E_0$, where
  $E_{\text{ad}_\text{p}}$ was obtained as $E_{\text{ad}_\text{H}}$ but using the constant,
  nuclear mass of the proton for $m_\text{p}$. \\
  $^\text{f}$ Experimental data not available. \\  
  $^\text{g}$ Note that the experimental uncertainty
  is an order-of-magnitude larger for these term values
  than for the others. \\
  $^\text{h}$ Neglect of delocalization to the inner well introduces an 
  at least 0.1~\cm\ error in the computed energy.
}  
\end{flushleft}

\clearpage

\clearpage
%


\begin{thebibliography}{38}
\expandafter\ifx\csname natexlab\endcsname\relax\def\natexlab#1{#1}\fi
\expandafter\ifx\csname bibnamefont\endcsname\relax
  \def\bibnamefont#1{#1}\fi
\expandafter\ifx\csname bibfnamefont\endcsname\relax
  \def\bibfnamefont#1{#1}\fi
\expandafter\ifx\csname citenamefont\endcsname\relax
  \def\citenamefont#1{#1}\fi
\expandafter\ifx\csname url\endcsname\relax
  \def\url#1{\texttt{#1}}\fi
\expandafter\ifx\csname urlprefix\endcsname\relax\def\urlprefix{URL }\fi
\providecommand{\bibinfo}[2]{#2}
\providecommand{\eprint}[2][]{\url{#2}}

\bibitem[{\citenamefont{M\'atyus and Teufel}(2019)}]{MaTe19}
\bibinfo{author}{\bibfnamefont{E.}~\bibnamefont{M\'atyus}} \bibnamefont{and}
  \bibinfo{author}{\bibfnamefont{S.}~\bibnamefont{Teufel}},
  \bibinfo{journal}{J. Chem. Phys.} \textbf{\bibinfo{volume}{151}},
  \bibinfo{pages}{014113} (\bibinfo{year}{2019}).

\bibitem[{\citenamefont{Teufel}(2003)}]{TeBook03}
\bibinfo{author}{\bibfnamefont{S.}~\bibnamefont{Teufel}},
  \emph{\bibinfo{title}{Adiabatic Perturbation Theory in Quantum Dynamics,
  Lecture Notes in Mathematics}} (\bibinfo{publisher}{Springer-Verlag},
  \bibinfo{address}{Berlin, Heidelberg, New York}, \bibinfo{year}{2003}).

\bibitem[{\citenamefont{Herman and Asgharian}(1966)}]{HeAs66}
\bibinfo{author}{\bibfnamefont{R.~M.} \bibnamefont{Herman}} \bibnamefont{and}
  \bibinfo{author}{\bibfnamefont{A.}~\bibnamefont{Asgharian}},
  \bibinfo{journal}{J. Mol. Spectrosc.} \textbf{\bibinfo{volume}{19}},
  \bibinfo{pages}{305} (\bibinfo{year}{1966}).

\bibitem[{\citenamefont{Herman and Ogilvie}(1998)}]{HeOg98}
\bibinfo{author}{\bibfnamefont{R.~M.} \bibnamefont{Herman}} \bibnamefont{and}
  \bibinfo{author}{\bibfnamefont{J.~F.} \bibnamefont{Ogilvie}},
  \bibinfo{journal}{Adv. Chem. Phys.} \textbf{\bibinfo{volume}{103}},
  \bibinfo{pages}{187} (\bibinfo{year}{1998}).

\bibitem[{\citenamefont{Schwenke}(2001{\natexlab{a}})}]{Sch01H2p}
\bibinfo{author}{\bibfnamefont{D.~W.} \bibnamefont{Schwenke}},
  \bibinfo{journal}{J. Chem. Phys.} \textbf{\bibinfo{volume}{114}},
  \bibinfo{pages}{1693} (\bibinfo{year}{2001}{\natexlab{a}}).

\bibitem[{\citenamefont{Pachucki and Komasa}(2009)}]{PaKo09}
\bibinfo{author}{\bibfnamefont{K.}~\bibnamefont{Pachucki}} \bibnamefont{and}
  \bibinfo{author}{\bibfnamefont{J.}~\bibnamefont{Komasa}},
  \bibinfo{journal}{J. Chem. Phys.} \textbf{\bibinfo{volume}{130}},
  \bibinfo{pages}{164113} (\bibinfo{year}{2009}).

\bibitem[{\citenamefont{Scherrer et~al.}(2017)\citenamefont{Scherrer, Agostini,
  Sebastiani, Gross, and Vuilleumier}}]{prx17}
\bibinfo{author}{\bibfnamefont{A.}~\bibnamefont{Scherrer}},
  \bibinfo{author}{\bibfnamefont{F.}~\bibnamefont{Agostini}},
  \bibinfo{author}{\bibfnamefont{D.}~\bibnamefont{Sebastiani}},
  \bibinfo{author}{\bibfnamefont{E.~K.~U.} \bibnamefont{Gross}},
  \bibnamefont{and}
  \bibinfo{author}{\bibfnamefont{R.}~\bibnamefont{Vuilleumier}},
  \bibinfo{journal}{Phys. Rev. X} \textbf{\bibinfo{volume}{7}},
  \bibinfo{pages}{031035} (\bibinfo{year}{2017}).

\bibitem[{\citenamefont{Bunker et~al.}(1977)\citenamefont{Bunker, McLarnon, and
  Moss}}]{BuMcMo77}
\bibinfo{author}{\bibfnamefont{P.~R.} \bibnamefont{Bunker}},
  \bibinfo{author}{\bibfnamefont{C.~J.} \bibnamefont{McLarnon}},
  \bibnamefont{and} \bibinfo{author}{\bibfnamefont{R.~E.} \bibnamefont{Moss}},
  \bibinfo{journal}{Mol. Phys.} \textbf{\bibinfo{volume}{33}},
  \bibinfo{pages}{425} (\bibinfo{year}{1977}).

\bibitem[{\citenamefont{Pachucki and Komasa}(2008)}]{PaKo08}
\bibinfo{author}{\bibfnamefont{K.}~\bibnamefont{Pachucki}} \bibnamefont{and}
  \bibinfo{author}{\bibfnamefont{J.}~\bibnamefont{Komasa}},
  \bibinfo{journal}{J. Chem. Phys.} \textbf{\bibinfo{volume}{129}},
  \bibinfo{pages}{034102} (\bibinfo{year}{2008}).

\bibitem[{\citenamefont{Mátyus}(2018{\natexlab{a}})}]{Ma18nonad}
\bibinfo{author}{\bibfnamefont{E.}~\bibnamefont{Mátyus}}, \bibinfo{journal}{J.
  Chem. Phys.} \textbf{\bibinfo{volume}{149}}, \bibinfo{pages}{194111}
  (\bibinfo{year}{2018}{\natexlab{a}}).

\bibitem[{\citenamefont{Mátyus}(2018{\natexlab{b}})}]{Ma18he2p}
\bibinfo{author}{\bibfnamefont{E.}~\bibnamefont{Mátyus}}, \bibinfo{journal}{J.
  Chem. Phys.} \textbf{\bibinfo{volume}{149}}, \bibinfo{pages}{194112}
  (\bibinfo{year}{2018}{\natexlab{b}}).

\bibitem[{\citenamefont{Przybytek et~al.}(2017)\citenamefont{Przybytek, Cencek,
  Jeziorski, and Szalewicz}}]{PrCeJeSz17}
\bibinfo{author}{\bibfnamefont{M.}~\bibnamefont{Przybytek}},
  \bibinfo{author}{\bibfnamefont{W.}~\bibnamefont{Cencek}},
  \bibinfo{author}{\bibfnamefont{B.}~\bibnamefont{Jeziorski}},
  \bibnamefont{and}
  \bibinfo{author}{\bibfnamefont{K.}~\bibnamefont{Szalewicz}},
  \bibinfo{journal}{Phys. Rev. Lett.} \textbf{\bibinfo{volume}{119}},
  \bibinfo{pages}{123401} (\bibinfo{year}{2017}).

\bibitem[{\citenamefont{Schwenke}(2001{\natexlab{b}})}]{Sch01H2O}
\bibinfo{author}{\bibfnamefont{D.~W.} \bibnamefont{Schwenke}},
  \bibinfo{journal}{J. Phys. Chem. A} \textbf{\bibinfo{volume}{105}},
  \bibinfo{pages}{2352} (\bibinfo{year}{2001}{\natexlab{b}}).

\bibitem[{\citenamefont{Bunker and Moss}(1977)}]{BuMo77}
\bibinfo{author}{\bibfnamefont{P.~R.} \bibnamefont{Bunker}} \bibnamefont{and}
  \bibinfo{author}{\bibfnamefont{R.~E.} \bibnamefont{Moss}},
  \bibinfo{journal}{Mol. Phys.} \textbf{\bibinfo{volume}{33}},
  \bibinfo{pages}{417} (\bibinfo{year}{1977}).

\bibitem[{\citenamefont{M\'atyus and Reiher}(2012)}]{MaRe12}
\bibinfo{author}{\bibfnamefont{E.}~\bibnamefont{M\'atyus}} \bibnamefont{and}
  \bibinfo{author}{\bibfnamefont{M.}~\bibnamefont{Reiher}},
  \bibinfo{journal}{J. Chem. Phys.} \textbf{\bibinfo{volume}{137}},
  \bibinfo{pages}{024104} (\bibinfo{year}{2012}).

\bibitem[{\citenamefont{Mátyus}(2019)}]{Ma19review}
\bibinfo{author}{\bibfnamefont{E.}~\bibnamefont{Mátyus}},
  \bibinfo{journal}{Mol. Phys.} \textbf{\bibinfo{volume}{117}},
  \bibinfo{pages}{590} (\bibinfo{year}{2019}).

\bibitem[{\citenamefont{Wolniewicz}(1998{\natexlab{a}})}]{Wo98rel}
\bibinfo{author}{\bibfnamefont{L.}~\bibnamefont{Wolniewicz}},
  \bibinfo{journal}{J. Chem. Phys.} \textbf{\bibinfo{volume}{109}},
  \bibinfo{pages}{2254} (\bibinfo{year}{1998}{\natexlab{a}}).

\bibitem[{\citenamefont{Light and {Carrington, Jr.}}(2000)}]{LiCa00}
\bibinfo{author}{\bibfnamefont{J.~C.} \bibnamefont{Light}} \bibnamefont{and}
  \bibinfo{author}{\bibfnamefont{T.}~\bibnamefont{{Carrington, Jr.}}},
  \bibinfo{journal}{Adv. Chem. Phys.} \textbf{\bibinfo{volume}{114}},
  \bibinfo{pages}{263} (\bibinfo{year}{2000}).

\bibitem[{\citenamefont{Wolniewicz}(1998{\natexlab{b}})}]{Wo98HH}
\bibinfo{author}{\bibfnamefont{L.}~\bibnamefont{Wolniewicz}},
  \bibinfo{journal}{J. Chem. Phys.} \textbf{\bibinfo{volume}{108}},
  \bibinfo{pages}{1499} (\bibinfo{year}{1998}{\natexlab{b}}).

\bibitem[{\citenamefont{Reinhold et~al.}(1999)\citenamefont{Reinhold,
  Hogervorst, Ubachs, and Wolniewicz}}]{ReHoUbWo99}
\bibinfo{author}{\bibfnamefont{E.}~\bibnamefont{Reinhold}},
  \bibinfo{author}{\bibfnamefont{W.}~\bibnamefont{Hogervorst}},
  \bibinfo{author}{\bibfnamefont{W.}~\bibnamefont{Ubachs}}, \bibnamefont{and}
  \bibinfo{author}{\bibfnamefont{L.}~\bibnamefont{Wolniewicz}},
  \bibinfo{journal}{Phys. Rev. A} \textbf{\bibinfo{volume}{60}},
  \bibinfo{pages}{1258} (\bibinfo{year}{1999}).

\bibitem[{\citenamefont{Andersson and Elander}(2004)}]{AnEl04}
\bibinfo{author}{\bibfnamefont{S.}~\bibnamefont{Andersson}} \bibnamefont{and}
  \bibinfo{author}{\bibfnamefont{N.}~\bibnamefont{Elander}},
  \bibinfo{journal}{Phys. Rev. A} \textbf{\bibinfo{volume}{69}},
  \bibinfo{pages}{052507} (\bibinfo{year}{2004}).

\bibitem[{\citenamefont{Ross et~al.}(2006)\citenamefont{Ross, Yoshinari, Ogi,
  and Tsukiyama}}]{RoYoOgTs06}
\bibinfo{author}{\bibfnamefont{S.~C.} \bibnamefont{Ross}},
  \bibinfo{author}{\bibfnamefont{T.}~\bibnamefont{Yoshinari}},
  \bibinfo{author}{\bibfnamefont{Y.}~\bibnamefont{Ogi}}, \bibnamefont{and}
  \bibinfo{author}{\bibfnamefont{K.}~\bibnamefont{Tsukiyama}},
  \bibinfo{journal}{J. Chem. Phys.} \textbf{\bibinfo{volume}{125}},
  \bibinfo{pages}{133205} (\bibinfo{year}{2006}).

\bibitem[{\citenamefont{Reinhold et~al.}(1997)\citenamefont{Reinhold,
  Hogervorst, and Ubachs}}]{ReHoUb97}
\bibinfo{author}{\bibfnamefont{E.}~\bibnamefont{Reinhold}},
  \bibinfo{author}{\bibfnamefont{W.}~\bibnamefont{Hogervorst}},
  \bibnamefont{and} \bibinfo{author}{\bibfnamefont{W.}~\bibnamefont{Ubachs}},
  \bibinfo{journal}{Phys. Rev. Lett.} \textbf{\bibinfo{volume}{78}},
  \bibinfo{pages}{2543} (\bibinfo{year}{1997}).

\bibitem[{som()}]{som}
\bibinfo{note}{The supplementary material contains the improved electronic
  energies for the outer well of the $H\bar{H}\ ^1\Sigma_\text{g}^+$ electronic
  state obtained in the present work.}

\bibitem[{FeM()}]{FeMa19a}
\bibinfo{note}{D. Ferenc and E. M\'atyus, Precise computation of rovibronic
  resonances of molecular hydrogen: $EF~^1\Sigma_\text{g}^+$ inner-well
  rotational states. arXiv:1904.08609}.

\bibitem[{\citenamefont{Korobov}(2018)}]{Ko18H2p}
\bibinfo{author}{\bibfnamefont{V.~I.} \bibnamefont{Korobov}},
  \bibinfo{journal}{Mol. Phys.} \textbf{\bibinfo{volume}{116}},
  \bibinfo{pages}{93} (\bibinfo{year}{2018}).

\bibitem[{\citenamefont{Yu and Dressler}(1994)}]{YuDr94}
\bibinfo{author}{\bibfnamefont{S.}~\bibnamefont{Yu}} \bibnamefont{and}
  \bibinfo{author}{\bibfnamefont{K.}~\bibnamefont{Dressler}},
  \bibinfo{journal}{J. Chem. Phys.} \textbf{\bibinfo{volume}{101}},
  \bibinfo{pages}{7692} (\bibinfo{year}{1994}).

\bibitem[{\citenamefont{Panati et~al.}(2007)\citenamefont{Panati, Spohn, and
  Teufel}}]{PaSpTe07}
\bibinfo{author}{\bibfnamefont{G.}~\bibnamefont{Panati}},
  \bibinfo{author}{\bibfnamefont{H.}~\bibnamefont{Spohn}}, \bibnamefont{and}
  \bibinfo{author}{\bibfnamefont{S.}~\bibnamefont{Teufel}},
  \bibinfo{journal}{ESAIM: Mathematical Modelling and Numerical Analysis}
  \textbf{\bibinfo{volume}{41}}, \bibinfo{pages}{297} (\bibinfo{year}{2007}).

\bibitem[{\citenamefont{Polyansky and Tennyson}(1999)}]{PoTe99}
\bibinfo{author}{\bibfnamefont{O.~L.} \bibnamefont{Polyansky}}
  \bibnamefont{and} \bibinfo{author}{\bibfnamefont{J.}~\bibnamefont{Tennyson}},
  \bibinfo{journal}{J. Chem. Phys.} \textbf{\bibinfo{volume}{110}},
  \bibinfo{pages}{5056} (\bibinfo{year}{1999}).

\bibitem[{\citenamefont{Diniz et~al.}(2013)\citenamefont{Diniz, Mohallem,
  Alijah, Pavanello, Adamowicz, Polyansky, and Tennyson}}]{DiMoAl13}
\bibinfo{author}{\bibfnamefont{L.~G.} \bibnamefont{Diniz}},
  \bibinfo{author}{\bibfnamefont{J.~R.} \bibnamefont{Mohallem}},
  \bibinfo{author}{\bibfnamefont{A.}~\bibnamefont{Alijah}},
  \bibinfo{author}{\bibfnamefont{M.}~\bibnamefont{Pavanello}},
  \bibinfo{author}{\bibfnamefont{L.}~\bibnamefont{Adamowicz}},
  \bibinfo{author}{\bibfnamefont{O.~L.} \bibnamefont{Polyansky}},
  \bibnamefont{and} \bibinfo{author}{\bibfnamefont{J.}~\bibnamefont{Tennyson}},
  \bibinfo{journal}{Phys. Rev. A} \textbf{\bibinfo{volume}{88}},
  \bibinfo{pages}{032506} (\bibinfo{year}{2013}).

\bibitem[{\citenamefont{Mátyus et~al.}(2014)\citenamefont{Mátyus,
  Szidarovszky, and Császár}}]{MaSzCs14}
\bibinfo{author}{\bibfnamefont{E.}~\bibnamefont{Mátyus}},
  \bibinfo{author}{\bibfnamefont{T.}~\bibnamefont{Szidarovszky}},
  \bibnamefont{and} \bibinfo{author}{\bibfnamefont{A.~G.}
  \bibnamefont{Császár}}, \bibinfo{journal}{J. Chem. Phys.}
  \textbf{\bibinfo{volume}{141}}, \bibinfo{pages}{154111}
  (\bibinfo{year}{2014}).

\bibitem[{\citenamefont{Wang and Yan}(2018)}]{WaYa18}
\bibinfo{author}{\bibfnamefont{L.~M.} \bibnamefont{Wang}} \bibnamefont{and}
  \bibinfo{author}{\bibfnamefont{Z.-C.} \bibnamefont{Yan}},
  \bibinfo{journal}{Phys. Rev. A} \textbf{\bibinfo{volume}{97}},
  \bibinfo{pages}{060501(R)} (\bibinfo{year}{2018}).

\bibitem[{\citenamefont{Puchalski et~al.}(2018)\citenamefont{Puchalski,
  Spyszkiewicz, Komasa, and Pachucki}}]{PuSpKoPa18}
\bibinfo{author}{\bibfnamefont{M.}~\bibnamefont{Puchalski}},
  \bibinfo{author}{\bibfnamefont{A.}~\bibnamefont{Spyszkiewicz}},
  \bibinfo{author}{\bibfnamefont{J.}~\bibnamefont{Komasa}}, \bibnamefont{and}
  \bibinfo{author}{\bibfnamefont{K.}~\bibnamefont{Pachucki}},
  \bibinfo{journal}{Phys. Rev. Lett.} \textbf{\bibinfo{volume}{121}},
  \bibinfo{pages}{073001} (\bibinfo{year}{2018}).

\bibitem[{\citenamefont{Wolniewicz}(1995{\natexlab{a}})}]{Wo95pi}
\bibinfo{author}{\bibfnamefont{L.}~\bibnamefont{Wolniewicz}},
  \bibinfo{journal}{J. Mol. Spectrosc.} \textbf{\bibinfo{volume}{174}},
  \bibinfo{pages}{132} (\bibinfo{year}{1995}{\natexlab{a}}).

\bibitem[{\citenamefont{Wolniewicz}(1995{\natexlab{b}})}]{Wo95delta}
\bibinfo{author}{\bibfnamefont{L.}~\bibnamefont{Wolniewicz}},
  \bibinfo{journal}{J. Mol. Spectrosc.} \textbf{\bibinfo{volume}{169}},
  \bibinfo{pages}{329} (\bibinfo{year}{1995}{\natexlab{b}}).

\bibitem[{\citenamefont{Wolniewicz and Dressler}(1994)}]{WoDr94}
\bibinfo{author}{\bibfnamefont{L.}~\bibnamefont{Wolniewicz}} \bibnamefont{and}
  \bibinfo{author}{\bibfnamefont{K.}~\bibnamefont{Dressler}},
  \bibinfo{journal}{J. Chem. Phys.} \textbf{\bibinfo{volume}{100}},
  \bibinfo{pages}{444} (\bibinfo{year}{1994}).

\bibitem[{\citenamefont{Wolniewicz et~al.}(1998)\citenamefont{Wolniewicz,
  Simbotin, and Dalgarno}}]{WoSiDa98}
\bibinfo{author}{\bibfnamefont{L.}~\bibnamefont{Wolniewicz}},
  \bibinfo{author}{\bibfnamefont{I.}~\bibnamefont{Simbotin}}, \bibnamefont{and}
  \bibinfo{author}{\bibfnamefont{A.}~\bibnamefont{Dalgarno}},
  \bibinfo{journal}{Astrophys. J. Suppl. Ser.} \textbf{\bibinfo{volume}{115}},
  \bibinfo{pages}{293} (\bibinfo{year}{1998}).

\bibitem[{\citenamefont{Beyer and Merkt}(2016)}]{BeMe16}
\bibinfo{author}{\bibfnamefont{M.}~\bibnamefont{Beyer}} \bibnamefont{and}
  \bibinfo{author}{\bibfnamefont{F.}~\bibnamefont{Merkt}}, \bibinfo{journal}{J.
  Mol. Spectrosc.} \textbf{\bibinfo{volume}{330}}, \bibinfo{pages}{147}
  (\bibinfo{year}{2016}).

\end{thebibliography}

\end{document}